# FERMILAB SWITCHYARD RESONANT BEAM POSITION MONITOR ELECTRONICS UPGRADE RESULTS*

T. Petersen†, J. Diamond, N. Liu, P. S. Prieto, D. Slimmer, A. Watts, Fermi National Accelerator Laboratory, Batavia, IL 60510, USA

*Abstract*

The readout electronics for the resonant beam position monitors (BPMs) in the Fermilab Switchyard (SY) have been upgraded, utilizing a low noise amplifier transition board and Fermilab designed digitizer boards. The stripline BPMs are estimated to have an average signal output of between -110 dBm and -80 dBm, with an estimated peak output of -70 dBm. The external resonant circuit is tuned to the SY machine frequency of 53.10348 MHz. Both the digitizer and transition boards have variable gain in order to accommodate the large dynamic range and irregularity of the resonant extraction spill. These BPMs will aid in auto-tuning of the SY beamline as well as enabling operators to monitor beam position through the spill.

## INTRODUCTION

The Fermilab Switchyard Stripline beam position monitors (BPMs) have been outfitted with new electronics. They consist of a transition board, VME processor (MVME 2401 or 5500), and an 8-channel, 125 MHz digitizer board. The transition boards condition the incoming signal before it arrives at the digitizers. This ensures that the signal is the appropriate magnitude to be read by the digitizer, as well as ensuring that only the frequency of interest is being detected. The large dynamic range and adjustable gain are the most considerable upgrades to the electronics system [1].

## SYSTEM OVERVIEW

### Resonant BPMs

The Switchyard BPMs have an external coil which in conjunction with the plate capacitance acts as a resonant circuit at 53 MHz. This increases the low intensity levels observed during slow spill. These signal levels are expected to be between -110 dBm to -80 dBm [2], with peak levels reaching higher due to the nature of the spill.

### Transition Board

The transition board electronics were designed with a large dynamic range in mind, with three 23 dB gain low noise amplifiers (LNAs) and a programmable attenuator with up to 31.5 dB of attenuation (in 0.5 dB steps) for an effective gain of 37.5-69 dB. This is shown in Figure 1. In addition the signal passes through a bandpass filter (BW of 5 MHz, centered at 53.10348 MHz) in order to increase the signal to noise ratio. This analog path also taps the signal after the first LNA to be read through a logarithmic amplifier, allowing the implementation of a control loop to determine the appropriate attenuation level for a given incoming RF power level. This is important so that the analog signal falls within the dynamic range of the Digitizer's ADC.

The digital portion of the transition board is controlled via a microcontroller. This is used to read in the incoming power level from the log amp, set the digital attenuator value, and communicate with the VME processor card through CAN bus protocol.

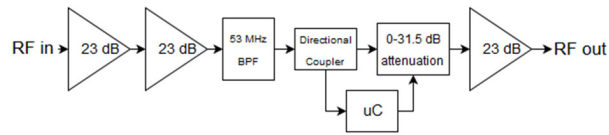

Figure 1: Block diagram of transition board.

### Digitizer and DSP Chain

The analog signal is read into an 8 channel digitizer with a sampling rate of 122.163 MHz. This was chosen to be above the Nyquist rate for the RF of interest (2.3 times the machine RF), and this clock signal is supplied via a PLL that is locked to the SY RF. The BPM pickup signal is digitally processed in an FPGA, with the DSP chain being detailed in Figure 2.

The ADC signal is down-converted from 53.10348 MHz to near base-band via mixing with an NCO produced signal, which produces an I&Q pair. This is done in order to get the signal close to DC, and still preserve phase data for magnitude calculation. When the NCO frequency was too close to the machine frequency of 53.10348 MHz, the down-converted frequency could be seen modulating the position readout. This was determined to be caused by truncation errors in the data chain, and was mitigated by choosing the down-converted frequency such that it could be easily notched out by the moving average filter.

Since the Switchyard slow spill lasts for approximately 4 seconds, the number of total samples for the spill needed to be reduced in order to effectively read out the data. The bulk of the sampling rate reduction was done by using a decimating CIC filter, with a decimation rate of 2048. The CIC filter has the benefit of decimation while also averaging the data. This is beneficial since the spill is not uniform, and the averaging assures that all power is accounted for in the accumulated intensity readout.

The following two FIR filters also have a decimation rate of 2, for a total decimation value of 8192 and final sampling frequency of 14.9 kHz. This is the rate at which the magnitude data (as seen in Figure 2) is reported. The intensity is calculated as the square root of the sum of squares of the two magnitudes. This value is reported separately from the individual magnitude data, which is further processed once it is passed to ACNET, Fermilab's Accelerator

---

* Work supported by the Fermi National Accelerator laboratory, operated by Fermi Research Alliance LLC, under contract No. DE-AC02-07CH11359 with the US Department of Energy
† tpeterse@fnal.gov

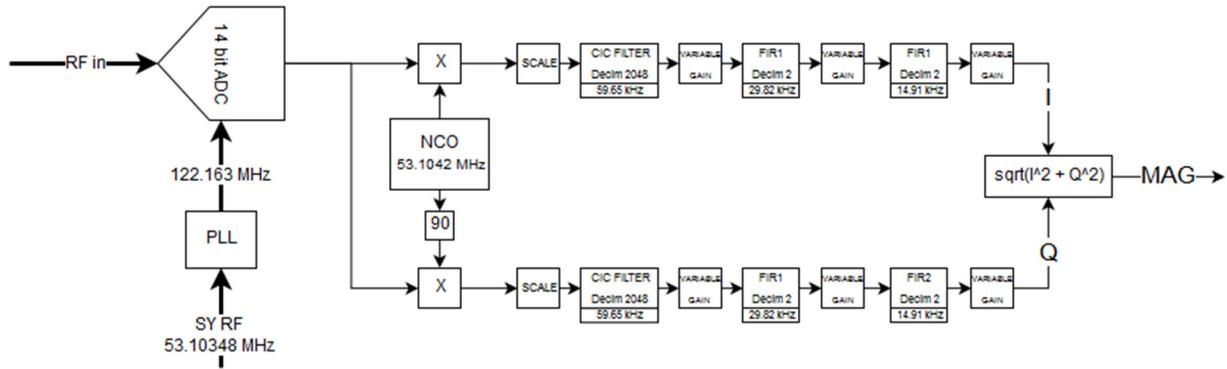

Figure 2: Block diagram of the DSP chain for the Switchyard Digitizers, for a single BPM plate.

Control system. Lastly, the magnitude data is passed through a moving average filter (with 160 points) before being reported to ACNET at 720 Hz.

Once in ACNET, the data for the magnitude read on each plate is used to calculate the position using a linear scale factor and offset.

## RESULTS AND FUTURE WORK

### Results

The new BPM electronics system has been installed and taking data for almost a year. Below, in Figure 3, is a plot of the beam position throughout a spill. The green trace is the calculated position, while the red and yellow traces are the magnitudes on either plate. Large spikes can be noticed throughout the spill, which illustrates the importance of making the system sensitive to the largest dynamic range possible.

Figure 4 shows the position (green) and accumulated intensity (red) throughout a spill. The intensity measurement still requires some fine tuning of the calibration in order to get a precise measurement, but the system has shown consistent measurements of both position and intensity.

### Usage

The majority of the use of the BPMs has been to begin an auto-tuning program for the Switchyard beamline. This uses an averaged position for each BPM to determine the best tuning. The real time data plotting capability of the system has also been advantageous, as it has helped diagnose an issue seen in the beamline known as beam roll (when the beam position moves in time throughout the spill). This issue and subsequent fix can be seen below, as illustrated in Figure 5.

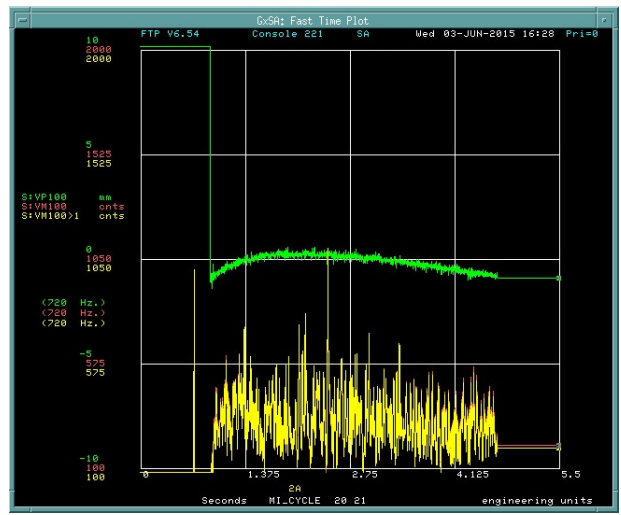

Figure 3: Calculated beam position (green) plotted throughout the spill, along with BPM plate intensities (red and yellow).

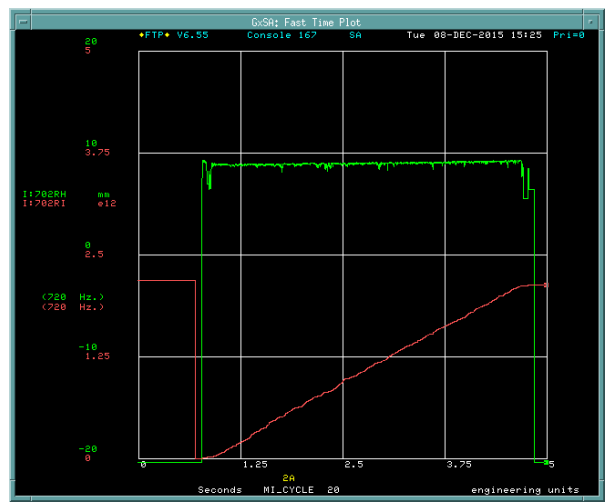

Figure 4: BPM position (green) plotted with reported accumulated intensity (red).

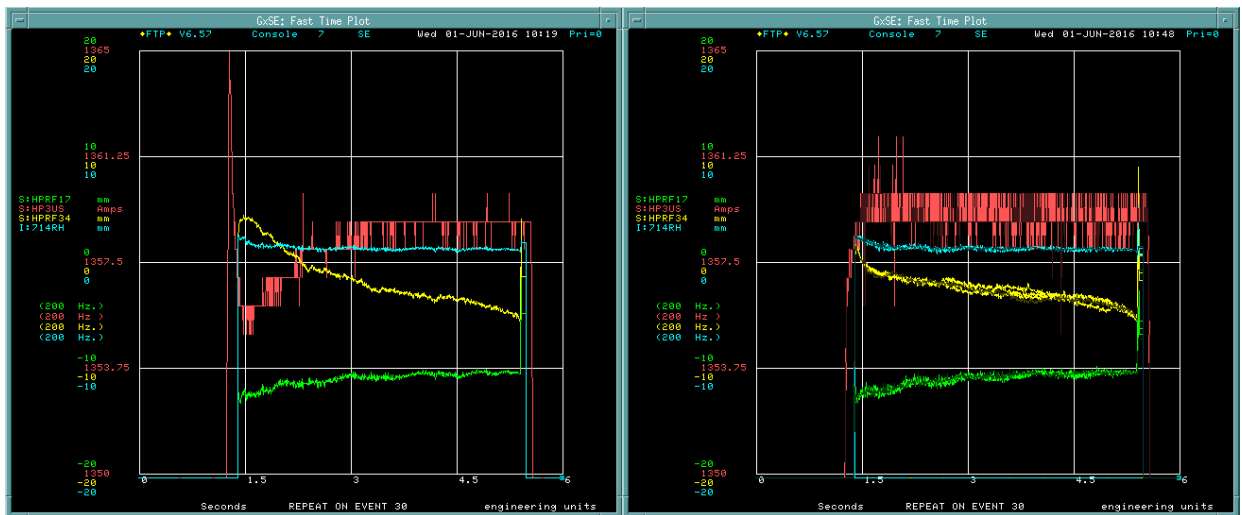

Figure 5: The plot on the left shows three beam positions as blue, green, and yellow (yellow being the one of interest) as well as power supply current (for a main dipole string) in red. It can be seen that as the power supply current is better regulated, the beam roll is much less prominent.

*Future Work*

Future work for the BPM system mostly entails finer calibration of the BPM position readout and calibration of the intensity readout. The intensity measurement has been the most difficult, as it proves difficult to make measurements on the resonant circuit without changing its properties. A separate approach has been taken, to begin calibrating a better understood device (resonant cavity spill monitor) in order to calibrate the BPMs off of that.

Additional work entails full commissioning of an adaptive system in which the log amp readings will set the attenuation value. Currently the system has been running at a static attenuation value, but as the intensity of the beam increases this will become a necessary feature. This will require a large lookup table that will set the appropriate scaling factor for each possible value, compensating for small differences in each channel's attenuation setting accuracy. Other work includes continuing the electronic replacement down the SY line.

## CONCLUSION

The new electronic system for the Fermilab Switchyard Beam Position Monitors has been installed and taking data. While some work remains, the BPMs have begun providing position measurements for the SY beamline.

## ACKNOWLEDGEMENTS

The authors would like to acknowledge the hard work and support from technicians, engineers, and physicists from the Instrumentation, Main Injector, and External Beams Departments.

## REFERENCES

[1] R. Fuja, *et. al*., Electronics Design for the Fermilab Switchyard Beam Position Monitor System.
[2] P. Stabile, *et. al*., Beam Position Monitor Electronics Upgrade for Fermilab Switchyard.